\documentclass[aps,prd,reprint,preprintnumbers,superscriptaddress,showpacs,twocolumn]{revtex4-1}
\usepackage{latexsym,graphicx,amssymb,amsmath,mathrsfs}
\usepackage{setspace,bm}
\usepackage[breaklinks, colorlinks=true, pdfstartview=FitV, linkcolor=red, citecolor=blue, urlcolor=blue]{hyperref}
\usepackage[usenames]{color}
\usepackage{latexsym}
\usepackage{epstopdf}
\usepackage{mathtools}
\usepackage{amssymb}

\usepackage[utf8]{inputenc}
\usepackage{tikz}
\usetikzlibrary{shapes,arrows}

\newcommand{\hyphen}{\mathchar`-}
\allowdisplaybreaks[1]
\usepackage[normalem]{ulem}  % \sout{old text} for strikeout

\begin{document}
\preprint{YITP-21-101}

\title{Gauge invariant input to neural network for path optimization method}

\author{Yusuke Namekawa}
\email[]{namekawa@yukawa.kyoto-u.ac.jp}
\affiliation{Yukawa Institute for Theoretical Physics, Kyoto University, Kyoto 606-8502, Japan}

\author{Kouji Kashiwa}
\affiliation{Fukuoka Institute of Technology, Wajiro, Fukuoka 811-0295, Japan}

\author{Akira Ohnishi}
\affiliation{Yukawa Institute for Theoretical Physics, Kyoto University, Kyoto 606-8502, Japan}

\author{Hayato Takase}
\noaffiliation{}

\begin{abstract}
We investigate the efficiency of a gauge invariant input to a neural network for the path optimization method.
While the path optimization with a completely gauge-fixed link-variable input has successfully tamed the sign problem in a simple gauge theory, the optimization does not work well when the gauge degrees of freedom remain.
We propose to employ a gauge invariant input, such as a plaquette, to overcome this problem.
The efficiency of the gauge invariant input to the neural network is evaluated for the 2-dimensional $U(1)$ gauge theory with a complex coupling.
The average phase factor is significantly enhanced by the path optimization with the plaquette input, indicating good control of the sign problem.
It opens a possibility that the path optimization is available to complicated gauge theories, including Quantum Chromodynamics, in a realistic setup.
\end{abstract}

\maketitle

\section{Introduction}
Exploring the phase structure of gauge theories at finite temperature ($T$) and chemical potential ($\mu$) is an interesting and important subject not only in particle and nuclear physics but also in astrophysics.
Quantitative understanding of heavy-ion experiments as well as the equation of state for neutron stars requires non-perturbative information of Quantum Chromodynamics (QCD) in the $T\hyphen\mu$ plane.
It is, however, a difficult task due to the sign problem.
Traditional Monte Carlo approaches work at low density, but fail in the middle and high density regions.

Recently, several new methods are developed to overcome the sign problem
at high densities, e.g. $\mu/T \geq 1$.
The complex Langevin method~\cite{Klauder:1983sp,Parisi:1984cs} is a stochastic quantization with complexified variables.
It is a non-Monte Carlo approach and thus free from the sign problem.
The low computational cost allows us to apply it to 4-dimensional QCD at finite density~\cite{Sexty:2013ica,Aarts:2014bwa,Fodor:2015doa,Nagata:2018mkb,Kogut:2019qmi,Sexty:2019vqx,Scherzer:2020kiu,Ito:2020mys}.
The tensor renormalization group method~\cite{Levin:2006jai} is a coarse graining algorithm using a tensor network.
It is also a non-Monte Carlo method.
Although the computational cost is extremely high, it has been vigorously tested even in 4-dimensional theoretical models~\cite{Akiyama:2019xzy,Akiyama:2020ntf,Akiyama:2020soe,Akiyama:2021zhf}.
Recent improved algorithms considerably reduce the cost~\cite{Kadoh:2019kqk,Kadoh:2021fri}.
The Lefschetz thimble method~\cite{Witten:2010cx} is a Monte Carlo scheme that complexifies variables and determines the integration path by solving an anti-holomorphic flow equation from fixed points such that the imaginary part of the action is constant.
Cauchy's integral theorem ensures the integral is independent of a choice of the integration path, if the path is given as a result of continuous deformation from the original path~\cite{Alexandru:2015sua}, crosses no poles, and the integral at infinity has no contribution.
The numerical study has been started with Langevin algorithm~\cite{Cristoforetti:2012su}, Metropolis algorithm~\cite{Mukherjee:2013aga}, and Hybrid Monte Carlo algorithm~\cite{Fujii:2013sra}.
The high computational cost is the main bottleneck of this method, but the algorithm development is overcoming it~\cite{Fukuma:2020fez,Fukuma:2021aoo}.
The path optimization method (POM)~\cite{Mori:2017pne,Mori:2017nwj}, also referred to as the sign-optimized manifold~\cite{Alexandru:2018fqp}, is an alternative approach that modifies the integration path by use of the machine learning via neural networks.
The machine learning finds the best path on which the sign problem is maximally weakened.
The POM successfully works for the complex $\lambda \phi^4$ theory~\cite{Mori:2017pne}, the Polyakov-loop extended Nambu--Jona-Lasinio model~\cite{Kashiwa:2018vxr,Kashiwa:2019lkv}, the Thirring model~\cite{Alexandru:2018fqp,Alexandru:2018ddf}, the $0+1$ dimensional bose gas~\cite{Bursa:2018ykf}, the $0+1$ dimensional QCD~\cite{Mori:2019tux}, the 2-dimensional $U(1)$ gauge theory with complexified coupling constant~\cite{Kashiwa:2020brj}, as well as noise reduction in observables~\cite{Detmold:2021ulb}.
The recent progress of the complexified path approaches is reviewed in Ref.~\cite{Alexandru:2020wrj}.

A key issue of the POM in gauge theories is control of the gauge degrees of freedom.
In $0+1$ dimensional QCD at finite density~\cite{Mori:2019tux}, the POM works with and without the gauge fixing.
In higher dimensions, however, the gauge fixing is required for the neural networks to find an improved path.
The effect of the gauge fixing is discussed in the 2-dimensional $U(1)$ gauge theory with complexified coupling constant~\cite{Kashiwa:2020brj}.
The average phase factor, an indicator of the sign problem, is never improved without the gauge fixing.
As we reduce the gauge degrees of freedom by the gauge fixing, the average phase factor is enhanced better.

Based on this result, we propose to adopt gauge invariant input for the optimization process.
The link variables are no longer direct input to the neural network.
We first construct a gauge invariant quantity and use it as the input.
We employ the simplest gauge invariant input, plaquette, in this study.
A similar idea is employed as a part of lattice gauge equivariant Convolutional Neural Networks~\cite{Favoni:2020reg}.
The performance of the POM with the gauge invariant input is demonstrated in the 2-dimensional $U(1)$ gauge theory with a complex coupling.
The sign problem originates from the imaginary part of the complex coupling.
The above-mentioned several methods have been tested for this theory~\cite{Kashiwa:2020brj,Pawlowski:2021bbu}.
Since the analytic result is available~\cite{Wiese:1988qz,Rusakov:1990rs,Bonati:2019ylr}, we can utilize it for verification of the simulation results.

This paper is organized as follows.
In the next section, we explain the formulation of the 2-dimensional lattice $U(1)$ gauge theory and the path optimization method.
Our numerical setup and results are presented in Sec.~\ref{Sec:Numerical}.
Section~\ref{Sec:Summary} summarizes this paper.

\section{Formulation}
\label{Sec:Formulation}

\subsection{$2$-dimensional $U(1)$ gauge action}

The gauge action is Wilson's plaquette action~\cite{Wilson:1974sk} given by
\begin{align}
  S_{\rm G} &= - \frac{\beta}{2} \sum_n \left( P_{n, 12} + P_{n, 12}^{-1} \right),
  \label{Eq:action_G}
\end{align}
where $n$ represents the lattice site, and $\beta = 1 / (g a)^2$ is an overall constant consisting of the gauge coupling constant $g$ and the lattice spacing $a$.
$P$ ($P^{-1}$) is the plaquette (its inverse). The definition is
\begin{align}
  P_{n, 12} := U_{n, 1} \, U_{n + \hat{1}, 2} \, U^{-1}_{n + \hat{2}, 1} \, U^{-1}_{n, 2},
\end{align}
where $\hat{\mu}$ is a unit vector in $\mu$-direction and
$U_{n, \mu}$ with $\mu = 1, 2$ are the $U(1)$ link variables.
We impose a periodic boundary condition in each direction.

In addition to the plaquette, we measure expectation values of the topological charge $Q$ defined on the lattice,
\begin{align}
  Q
  &:= - \frac{i}{4 \pi} \sum_n (P_{n, 12} - P_{n, 12}^{-1}).
  \label{Eq:Q}
\end{align}
In the continuum limit, $Q$ recovers the continuum form $Q = (1 / 4 \pi) \int d^2 x \,  \epsilon^{\mu\nu} F_{\mu\nu}(x)$.

The analytic result of this theory has been obtained~\cite{Wiese:1988qz,Rusakov:1990rs,Bonati:2019ylr}.
The partition function $Z$ is represented through the modified Bessel function $I_n(\beta)$,
\begin{align}
  Z
  &= \sum_{n = - \infty}^{+ \infty} I_n(\beta)^V,
  \label{Eq:exact}
  \\
  I_n(\beta)
  &:= \frac{1}{2 \pi} \int_{- \pi}^{\pi} d\phi \, e^{\beta \cos \phi - i n \phi},
\end{align}
where $V = N_1 N_2$ is the volume factor with $N_\mu$ being the lattice size in $\mu$-direction.
Since $I_n(\beta)$ is well-defined for all complex values of $\beta$, the analytic solution is available over the whole domain of $\beta$.

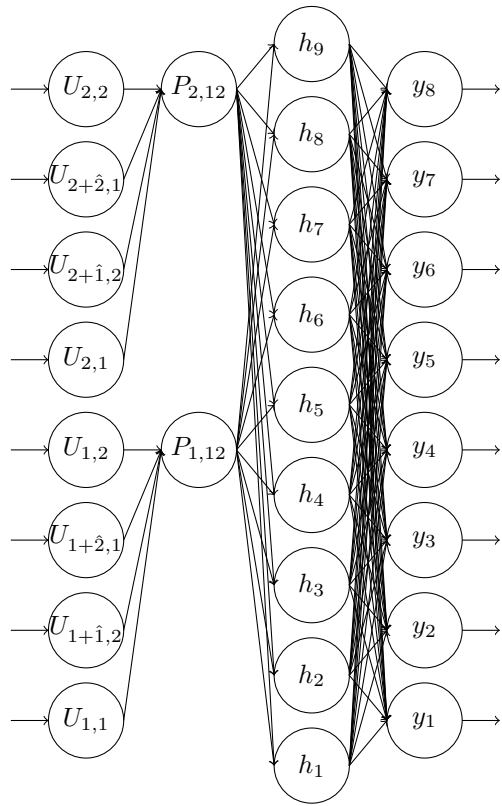
\begin{figure}[t]
 \centering
 \begin{tikzpicture}
  \draw[->] (0.0, 0.6 + 1.2 * 7)--(0.5, 0.6 + 1.2 * 7);
  \draw[->] (0.0, 0.6 + 1.2 * 6)--(0.5, 0.6 + 1.2 * 6);
  \draw[->] (0.0, 0.6 + 1.2 * 5)--(0.5, 0.6 + 1.2 * 5);
  \draw[->] (0.0, 0.6 + 1.2 * 4)--(0.5, 0.6 + 1.2 * 4);
  \draw[->] (0.0, 0.6 + 1.2 * 3)--(0.5, 0.6 + 1.2 * 3);
  \draw[->] (0.0, 0.6 + 1.2 * 2)--(0.5, 0.6 + 1.2 * 2);
  \draw[->] (0.0, 0.6 + 1.2 * 1)--(0.5, 0.6 + 1.2 * 1);
  \draw[->] (0.0, 0.6 + 1.2 * 0)--(0.5, 0.6 + 1.2 * 0);
  \draw(1.0, 0.6 + 1.2 * 7) circle (0.5) node{$U_{2, 2}$};
  \draw(1.0, 0.6 + 1.2 * 6) circle (0.5) node{$U_{2 + \hat{2}, 1}$};
  \draw(1.0, 0.6 + 1.2 * 5) circle (0.5) node{$U_{2 + \hat{1}, 2}$};
  \draw(1.0, 0.6 + 1.2 * 4) circle (0.5) node{$U_{2, 1}$};
  \draw(1.0, 0.6 + 1.2 * 3) circle (0.5) node{$U_{1, 2}$};
  \draw(1.0, 0.6 + 1.2 * 2) circle (0.5) node{$U_{1 + \hat{2}, 1}$};
  \draw(1.0, 0.6 + 1.2 * 1) circle (0.5) node{$U_{1 + \hat{1}, 2}$};
  \draw(1.0, 0.6 + 1.2 * 0) circle (0.5) node{$U_{1, 1}$};
  \draw[->] (1.5, 0.6 + 1.2 * 7)--(2.0, 0.6 + 1.2 * 7);
  \draw[->] (1.5, 0.6 + 1.2 * 6)--(2.0, 0.6 + 1.2 * 7);
  \draw[->] (1.5, 0.6 + 1.2 * 5)--(2.0, 0.6 + 1.2 * 7);
  \draw[->] (1.5, 0.6 + 1.2 * 4)--(2.0, 0.6 + 1.2 * 7);
  \draw[->] (1.5, 0.6 + 1.2 * 3)--(2.0, 0.6 + 1.2 * 3);
  \draw[->] (1.5, 0.6 + 1.2 * 2)--(2.0, 0.6 + 1.2 * 3);
  \draw[->] (1.5, 0.6 + 1.2 * 1)--(2.0, 0.6 + 1.2 * 3);
  \draw[->] (1.5, 0.6 + 1.2 * 0)--(2.0, 0.6 + 1.2 * 3);
  \draw(2.5, 0.6 + 1.2 * 7) circle (0.5) node{$P_{2, 12}$};
  \draw(2.5, 0.6 + 1.2 * 3) circle (0.5) node{$P_{1, 12}$};
  \draw[->] (3.0, 0.6 + 1.2 * 7)--(3.5, 0.0 + 1.2 * 8);
  \draw[->] (3.0, 0.6 + 1.2 * 7)--(3.5, 0.0 + 1.2 * 7);
  \draw[->] (3.0, 0.6 + 1.2 * 7)--(3.5, 0.0 + 1.2 * 6);
  \draw[->] (3.0, 0.6 + 1.2 * 7)--(3.5, 0.0 + 1.2 * 5);
  \draw[->] (3.0, 0.6 + 1.2 * 7)--(3.5, 0.0 + 1.2 * 4);
  \draw[->] (3.0, 0.6 + 1.2 * 7)--(3.5, 0.0 + 1.2 * 3);
  \draw[->] (3.0, 0.6 + 1.2 * 7)--(3.5, 0.0 + 1.2 * 2);
  \draw[->] (3.0, 0.6 + 1.2 * 7)--(3.5, 0.0 + 1.2 * 1);
  \draw[->] (3.0, 0.6 + 1.2 * 7)--(3.5, 0.0 + 1.2 * 0);
  \draw[->] (3.0, 0.6 + 1.2 * 3)--(3.5, 0.0 + 1.2 * 8);
  \draw[->] (3.0, 0.6 + 1.2 * 3)--(3.5, 0.0 + 1.2 * 7);
  \draw[->] (3.0, 0.6 + 1.2 * 3)--(3.5, 0.0 + 1.2 * 6);
  \draw[->] (3.0, 0.6 + 1.2 * 3)--(3.5, 0.0 + 1.2 * 5);
  \draw[->] (3.0, 0.6 + 1.2 * 3)--(3.5, 0.0 + 1.2 * 4);
  \draw[->] (3.0, 0.6 + 1.2 * 3)--(3.5, 0.0 + 1.2 * 3);
  \draw[->] (3.0, 0.6 + 1.2 * 3)--(3.5, 0.0 + 1.2 * 2);
  \draw[->] (3.0, 0.6 + 1.2 * 3)--(3.5, 0.0 + 1.2 * 1);
  \draw[->] (3.0, 0.6 + 1.2 * 3)--(3.5, 0.0 + 1.2 * 0);
  \draw(4.0, 0.0 + 1.2 * 8) circle (0.5) node{$h_9$};
  \draw(4.0, 0.0 + 1.2 * 7) circle (0.5) node{$h_8$};
  \draw(4.0, 0.0 + 1.2 * 6) circle (0.5) node{$h_7$};
  \draw(4.0, 0.0 + 1.2 * 5) circle (0.5) node{$h_6$};
  \draw(4.0, 0.0 + 1.2 * 4) circle (0.5) node{$h_5$};
  \draw(4.0, 0.0 + 1.2 * 3) circle (0.5) node{$h_4$};
  \draw(4.0, 0.0 + 1.2 * 2) circle (0.5) node{$h_3$};
  \draw(4.0, 0.0 + 1.2 * 1) circle (0.5) node{$h_2$};
  \draw(4.0, 0.0 + 1.2 * 0) circle (0.5) node{$h_1$};
  \draw[->] (4.5, 0.0 + 1.2 * 8)--(5.0, 0.6 + 1.2 * 7);
  \draw[->] (4.5, 0.0 + 1.2 * 8)--(5.0, 0.6 + 1.2 * 6);
  \draw[->] (4.5, 0.0 + 1.2 * 8)--(5.0, 0.6 + 1.2 * 5);
  \draw[->] (4.5, 0.0 + 1.2 * 8)--(5.0, 0.6 + 1.2 * 4);
  \draw[->] (4.5, 0.0 + 1.2 * 8)--(5.0, 0.6 + 1.2 * 3);
  \draw[->] (4.5, 0.0 + 1.2 * 8)--(5.0, 0.6 + 1.2 * 2);
  \draw[->] (4.5, 0.0 + 1.2 * 8)--(5.0, 0.6 + 1.2 * 1);
  \draw[->] (4.5, 0.0 + 1.2 * 8)--(5.0, 0.6 + 1.2 * 0);
  \draw[->] (4.5, 0.0 + 1.2 * 7)--(5.0, 0.6 + 1.2 * 7);
  \draw[->] (4.5, 0.0 + 1.2 * 7)--(5.0, 0.6 + 1.2 * 6);
  \draw[->] (4.5, 0.0 + 1.2 * 7)--(5.0, 0.6 + 1.2 * 5);
  \draw[->] (4.5, 0.0 + 1.2 * 7)--(5.0, 0.6 + 1.2 * 4);
  \draw[->] (4.5, 0.0 + 1.2 * 7)--(5.0, 0.6 + 1.2 * 3);
  \draw[->] (4.5, 0.0 + 1.2 * 7)--(5.0, 0.6 + 1.2 * 2);
  \draw[->] (4.5, 0.0 + 1.2 * 7)--(5.0, 0.6 + 1.2 * 1);
  \draw[->] (4.5, 0.0 + 1.2 * 7)--(5.0, 0.6 + 1.2 * 0);
  \draw[->] (4.5, 0.0 + 1.2 * 6)--(5.0, 0.6 + 1.2 * 7);
  \draw[->] (4.5, 0.0 + 1.2 * 6)--(5.0, 0.6 + 1.2 * 6);
  \draw[->] (4.5, 0.0 + 1.2 * 6)--(5.0, 0.6 + 1.2 * 5);
  \draw[->] (4.5, 0.0 + 1.2 * 6)--(5.0, 0.6 + 1.2 * 4);
  \draw[->] (4.5, 0.0 + 1.2 * 6)--(5.0, 0.6 + 1.2 * 3);
  \draw[->] (4.5, 0.0 + 1.2 * 6)--(5.0, 0.6 + 1.2 * 2);
  \draw[->] (4.5, 0.0 + 1.2 * 6)--(5.0, 0.6 + 1.2 * 1);
  \draw[->] (4.5, 0.0 + 1.2 * 6)--(5.0, 0.6 + 1.2 * 0);
  \draw[->] (4.5, 0.0 + 1.2 * 5)--(5.0, 0.6 + 1.2 * 7);
  \draw[->] (4.5, 0.0 + 1.2 * 5)--(5.0, 0.6 + 1.2 * 6);
  \draw[->] (4.5, 0.0 + 1.2 * 5)--(5.0, 0.6 + 1.2 * 5);
  \draw[->] (4.5, 0.0 + 1.2 * 5)--(5.0, 0.6 + 1.2 * 4);
  \draw[->] (4.5, 0.0 + 1.2 * 5)--(5.0, 0.6 + 1.2 * 3);
  \draw[->] (4.5, 0.0 + 1.2 * 5)--(5.0, 0.6 + 1.2 * 2);
  \draw[->] (4.5, 0.0 + 1.2 * 5)--(5.0, 0.6 + 1.2 * 1);
  \draw[->] (4.5, 0.0 + 1.2 * 5)--(5.0, 0.6 + 1.2 * 0);
  \draw[->] (4.5, 0.0 + 1.2 * 4)--(5.0, 0.6 + 1.2 * 7);
  \draw[->] (4.5, 0.0 + 1.2 * 4)--(5.0, 0.6 + 1.2 * 6);
  \draw[->] (4.5, 0.0 + 1.2 * 4)--(5.0, 0.6 + 1.2 * 5);
  \draw[->] (4.5, 0.0 + 1.2 * 4)--(5.0, 0.6 + 1.2 * 4);
  \draw[->] (4.5, 0.0 + 1.2 * 4)--(5.0, 0.6 + 1.2 * 3);
  \draw[->] (4.5, 0.0 + 1.2 * 4)--(5.0, 0.6 + 1.2 * 2);
  \draw[->] (4.5, 0.0 + 1.2 * 4)--(5.0, 0.6 + 1.2 * 1);
  \draw[->] (4.5, 0.0 + 1.2 * 4)--(5.0, 0.6 + 1.2 * 0);
  \draw[->] (4.5, 0.0 + 1.2 * 3)--(5.0, 0.6 + 1.2 * 7);
  \draw[->] (4.5, 0.0 + 1.2 * 3)--(5.0, 0.6 + 1.2 * 6);
  \draw[->] (4.5, 0.0 + 1.2 * 3)--(5.0, 0.6 + 1.2 * 5);
  \draw[->] (4.5, 0.0 + 1.2 * 3)--(5.0, 0.6 + 1.2 * 4);
  \draw[->] (4.5, 0.0 + 1.2 * 3)--(5.0, 0.6 + 1.2 * 3);
  \draw[->] (4.5, 0.0 + 1.2 * 3)--(5.0, 0.6 + 1.2 * 2);
  \draw[->] (4.5, 0.0 + 1.2 * 3)--(5.0, 0.6 + 1.2 * 1);
  \draw[->] (4.5, 0.0 + 1.2 * 3)--(5.0, 0.6 + 1.2 * 0);
  \draw[->] (4.5, 0.0 + 1.2 * 2)--(5.0, 0.6 + 1.2 * 7);
  \draw[->] (4.5, 0.0 + 1.2 * 2)--(5.0, 0.6 + 1.2 * 6);
  \draw[->] (4.5, 0.0 + 1.2 * 2)--(5.0, 0.6 + 1.2 * 5);
  \draw[->] (4.5, 0.0 + 1.2 * 2)--(5.0, 0.6 + 1.2 * 4);
  \draw[->] (4.5, 0.0 + 1.2 * 2)--(5.0, 0.6 + 1.2 * 3);
  \draw[->] (4.5, 0.0 + 1.2 * 2)--(5.0, 0.6 + 1.2 * 2);
  \draw[->] (4.5, 0.0 + 1.2 * 2)--(5.0, 0.6 + 1.2 * 1);
  \draw[->] (4.5, 0.0 + 1.2 * 2)--(5.0, 0.6 + 1.2 * 0);
  \draw[->] (4.5, 0.0 + 1.2 * 1)--(5.0, 0.6 + 1.2 * 7);
  \draw[->] (4.5, 0.0 + 1.2 * 1)--(5.0, 0.6 + 1.2 * 6);
  \draw[->] (4.5, 0.0 + 1.2 * 1)--(5.0, 0.6 + 1.2 * 5);
  \draw[->] (4.5, 0.0 + 1.2 * 1)--(5.0, 0.6 + 1.2 * 4);
  \draw[->] (4.5, 0.0 + 1.2 * 1)--(5.0, 0.6 + 1.2 * 3);
  \draw[->] (4.5, 0.0 + 1.2 * 1)--(5.0, 0.6 + 1.2 * 2);
  \draw[->] (4.5, 0.0 + 1.2 * 1)--(5.0, 0.6 + 1.2 * 1);
  \draw[->] (4.5, 0.0 + 1.2 * 1)--(5.0, 0.6 + 1.2 * 0);
  \draw[->] (4.5, 0.0 + 1.2 * 0)--(5.0, 0.6 + 1.2 * 7);
  \draw[->] (4.5, 0.0 + 1.2 * 0)--(5.0, 0.6 + 1.2 * 6);
  \draw[->] (4.5, 0.0 + 1.2 * 0)--(5.0, 0.6 + 1.2 * 5);
  \draw[->] (4.5, 0.0 + 1.2 * 0)--(5.0, 0.6 + 1.2 * 4);
  \draw[->] (4.5, 0.0 + 1.2 * 0)--(5.0, 0.6 + 1.2 * 3);
  \draw[->] (4.5, 0.0 + 1.2 * 0)--(5.0, 0.6 + 1.2 * 2);
  \draw[->] (4.5, 0.0 + 1.2 * 0)--(5.0, 0.6 + 1.2 * 1);
  \draw[->] (4.5, 0.0 + 1.2 * 0)--(5.0, 0.6 + 1.2 * 0);
  \draw(5.5, 0.6 + 1.2 * 7) circle (0.5) node{$y_8$};
  \draw(5.5, 0.6 + 1.2 * 6) circle (0.5) node{$y_7$};
  \draw(5.5, 0.6 + 1.2 * 5) circle (0.5) node{$y_6$};
  \draw(5.5, 0.6 + 1.2 * 4) circle (0.5) node{$y_5$};
  \draw(5.5, 0.6 + 1.2 * 3) circle (0.5) node{$y_4$};
  \draw(5.5, 0.6 + 1.2 * 2) circle (0.5) node{$y_3$};
  \draw(5.5, 0.6 + 1.2 * 1) circle (0.5) node{$y_2$};
  \draw(5.5, 0.6 + 1.2 * 0) circle (0.5) node{$y_1$};
  \draw[->] (6.0, 0.6 + 1.2 * 7)--(6.5, 0.6 + 1.2 * 7);
  \draw[->] (6.0, 0.6 + 1.2 * 6)--(6.5, 0.6 + 1.2 * 6);
  \draw[->] (6.0, 0.6 + 1.2 * 5)--(6.5, 0.6 + 1.2 * 5);
  \draw[->] (6.0, 0.6 + 1.2 * 4)--(6.5, 0.6 + 1.2 * 4);
  \draw[->] (6.0, 0.6 + 1.2 * 3)--(6.5, 0.6 + 1.2 * 3);
  \draw[->] (6.0, 0.6 + 1.2 * 2)--(6.5, 0.6 + 1.2 * 2);
  \draw[->] (6.0, 0.6 + 1.2 * 1)--(6.5, 0.6 + 1.2 * 1);
  \draw[->] (6.0, 0.6 + 1.2 * 0)--(6.5, 0.6 + 1.2 * 0);
 \end{tikzpicture}
 \caption{Schematic picture of the neural network with the plaquette input. The link variable $U_{n, \mu}$ is converted to the plaquette $P_{n, \mu\nu}$ in the input layer. $h_i$ and $y_i$ are hidden and output layers, respectively.}
 \label{Fig:NN}
\end{figure}

\subsection{Path optimization method with plaquette and link input}
The path optimization method utilizes complexified dynamical variables to tame the sign problem.
In the case of the $U(1)$ gauge theory, the plaquette and the link variable are extended as
\begin{align}
  {\cal P}_{n,12}
  &= {\cal U}_{n,1} \, {\cal U}_{n+\hat{1},2} \, {\cal U}^{-1}_{n+\hat{2},1} \, {\cal U}^{-1}_{n,2},
  \\
  {\cal U}_{n,\mu}
  &= e^{ig {\cal A}_\mu(n+\hat{\mu}/2)} =: U_{n,\mu} \, e^{- y_n},
\end{align}
where ${\cal A}_\mu \in \mathbb{C}$.
The modification of the integral path is represented by $y_n \in \mathbb{R}$, originated from the imaginary part of ${\cal A}_\mu$, which is evaluated by a neural network consisting of the input, a single hidden, and the output layers, as follows.
The neural network is a mathematical model inspired by a brain, which is often used for the machine learning~\cite{McCulloch:1943logical,Rosenblatt:1958perceptron,Hebb2002::organization,Hinton2006:reducing}.
A sufficient number of the hidden layer units with a non-linear function called an activation function reproduce any continuous function, as proven by the universal approximation theorem~\cite{Cybenko:1989,Hornik:1991251}.
The variables in the hidden layer nodes ($h_j$) and the output  ($y_n$) are set as
\begin{align}
 h_j &= F(w^{(1)}_{ji} t_i + b_j),
 \label{Eq:FNN1}
 \\
 y_n &= \omega_n F(w^{(2)}_{nj} h_j + b_j),
 \label{Eq:FNN2}
\end{align}
where $i, j = 1, \cdots, 2 \times n_\mathrm{deg}$ with the number of the degree of freedom $n_\mathrm{deg}$ and $w$, $b$ and $\omega$ are parameters of the neural network.
An activation function $F$ defines a relation of data between two layers, the input and hidden layers as well as the hidden and output layers.
The activation function is taken to be a tangent hyperbolic function in this work.
Our choice of the input $t = \{ t_i \}$ is the plaquette or the link variable.
\begin{align}
 ({\rm i})  \, &t = {\rm Re } \, P, \, {\rm Im } \, P
 \label{Eq:t_P}
 \\
 ({\rm ii}) \, &t = {\rm Re } \, U, \, {\rm Im } \, U.
 \label{Eq:t_U}
\end{align}
We compare (i) with (ii) in terms of the efficiency of the neural network.
A schematic picture of our neural network with the plaquette input is given in Fig.~\ref{Fig:NN}.

The expectation value of a complexified observable ${\mathcal O}$ is calculated by
\begin{align}
  \langle {\mathcal O} \rangle
  &:= \frac{1}{Z} \int \mathcal{D} U {\mathcal O}(U) e^{- S_{\rm G}(U)}
  = \frac{1}{Z} \int_{\mathcal{C}}\, \mathcal{DU} \, {\mathcal O}(\mathcal{U}) \, e^{-S_{\rm G}(\mathcal{U})}
  \nonumber \\
  &= \frac{1}{Z} \int \mathcal{D}U\, \left[{\mathcal O}
  e^{i \theta} |J \, e^{-S_{\rm G}}|\right]_{\mathcal{U} \in \mathcal{C}}
  \nonumber \\
  &= 
  \frac{\langle {\mathcal O} \, e^{i \theta} \rangle_\mathrm{pq}}
          {\langle e^{i \theta} \rangle_\mathrm{pq}},
  \,
  \langle {\mathcal O} \rangle_\mathrm{pq}
  := \frac{1}{Z} \int \mathcal{D}U\, \left[{\mathcal O}
                 |J \, e^{-S_{\rm G}}|\right]_{\mathcal{U} \in \mathcal{C}},
\label{Eq:pq}
\end{align}
where
$\mathcal{C}$ is the integration path that specifies the complexified link variables
$\mathcal{U}=\mathcal{U}(U)$ and the Jacobian $J(\mathcal{U}(U))=\mathrm{det}(\partial \mathcal{U}/\partial U)$.
$\theta$ is the phase of $J \, e^{-S_{\rm G}} = e^{i \theta} |J \, e^{-S_{\rm G}}|$.
$\langle \cdots \rangle_\mathrm{pq}$ denotes the expectation value with the phase quenched Boltzmann weight.
In contrast to the naive reweighting, Eq.~\eqref{Eq:pq} is evaluated on the modified integration path where the sign problem is maximally weakened by the machine learning without change of the expectation value guaranteed by Cauchy's theorem.
This is the advantage of the path optimization method.

The cost function controls optimization through the neural network.
We apply the following cost function ${\cal F}$ to minimize the sign problem,
\begin{align}
  {\cal F}[y(t)]
  = |Z| \left( | \langle e^{i \theta(t)} \rangle_\mathrm{pq} |^{-1} - 1 \right).
  \label{Eq:cost_func}
\end{align}
We evaluate it by the exponential moving average (EMA) as in Ref.~\cite{Kashiwa:2020brj}.
Using this cost function~\eqref{Eq:cost_func}, the neural network finds the best path that enhances $e^{i \theta(t)}$ as much as possible.

\begin{figure}[t]
 \centering
 \includegraphics[width=7.5cm]{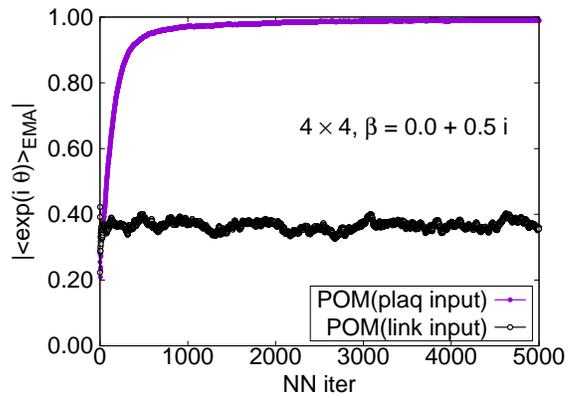}
 \caption{Neural network iteration dependence of the exponential moving average phase factors at $\beta = 0.0 + 0.5 i$ on $4 \times 4$ lattice.}
 \label{Fig:iter-apf_beta_i}
\end{figure}

\begin{figure}[ht]
 \centering
 \includegraphics[width=7.5cm]{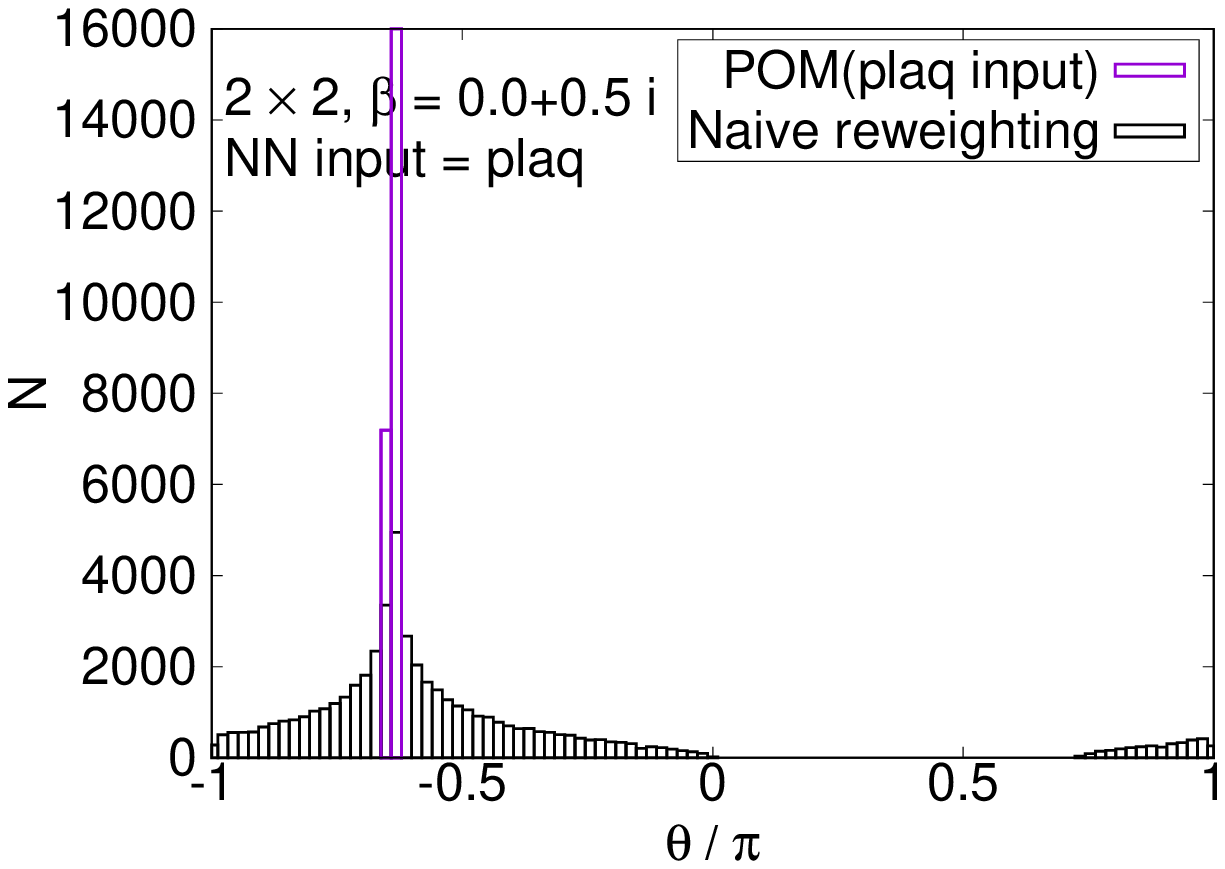}
 \includegraphics[width=7.5cm]{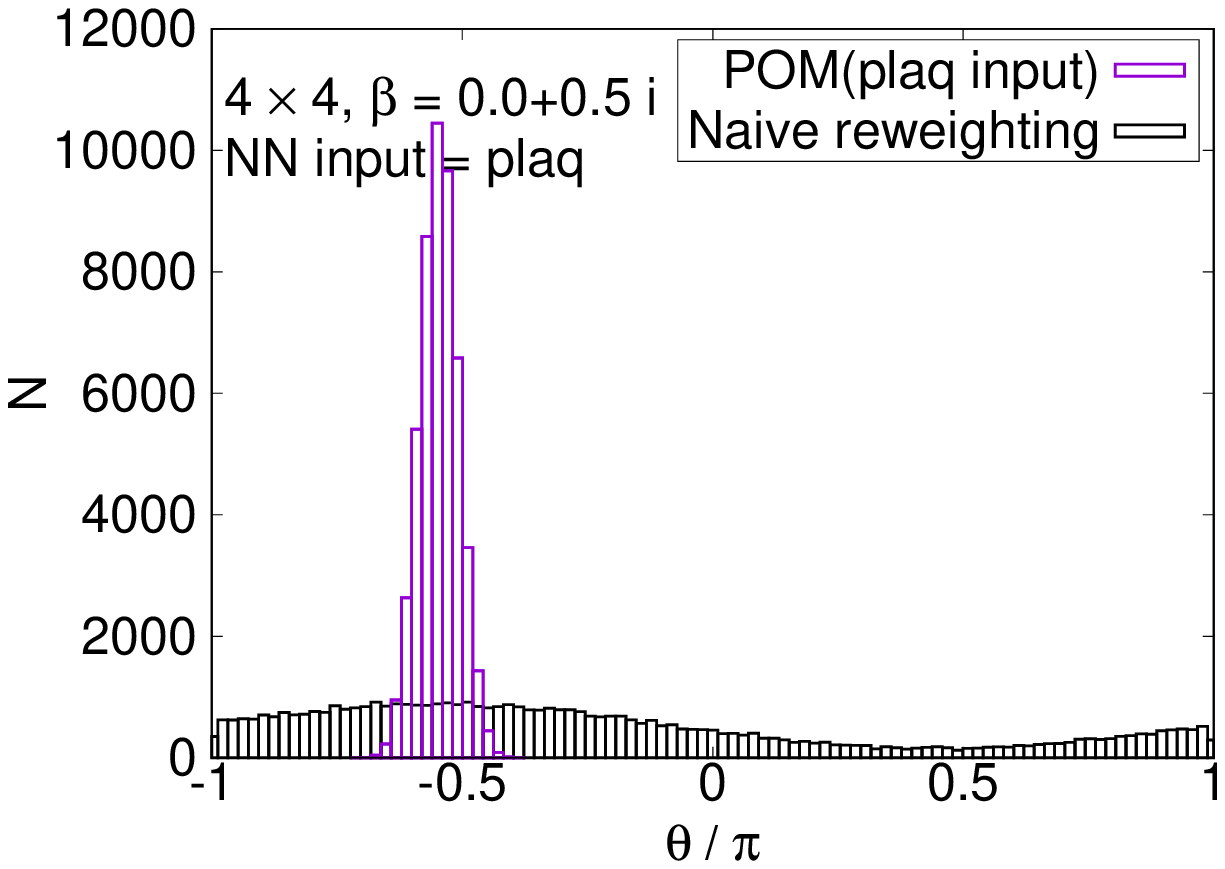}
 \includegraphics[width=7.5cm]{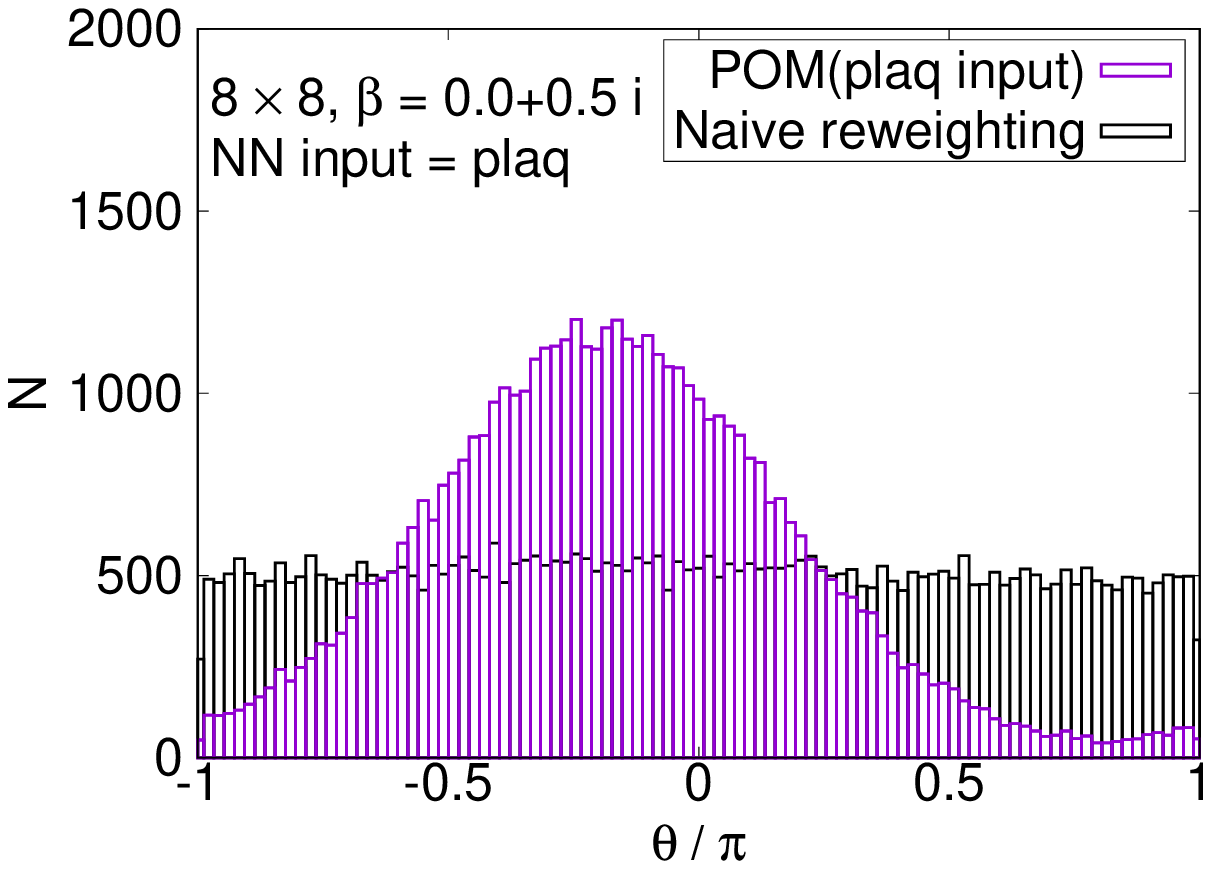}
 \caption{Histogram of the phases with and without the path optimization at $\beta = 0.0 + 0.5 i$ on $2 \times 2$ (upper panel), $4 \times 4$ (middle panel), and $8 \times 8$ (lower panel) lattices.}
 \label{Fig:histogram_beta0.5}
\end{figure}

\begin{figure}[ht]
 \centering
 \includegraphics[width=7.5cm]{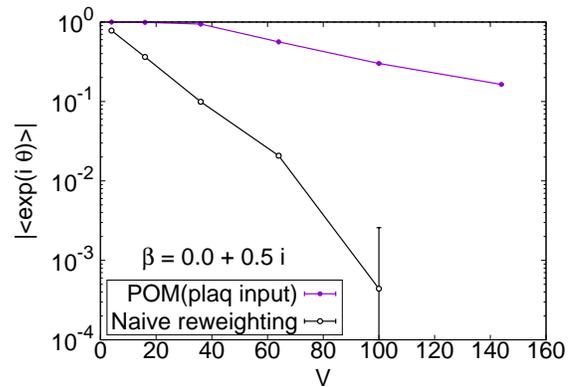}
 \caption{Volume dependence of the average phase factor at $\beta = 0.0 + 0.5 i$.}
 \label{Fig:L2-apf_beta}
\end{figure}

\begin{figure}[ht]
 \centering
 \includegraphics[width=7.5cm]{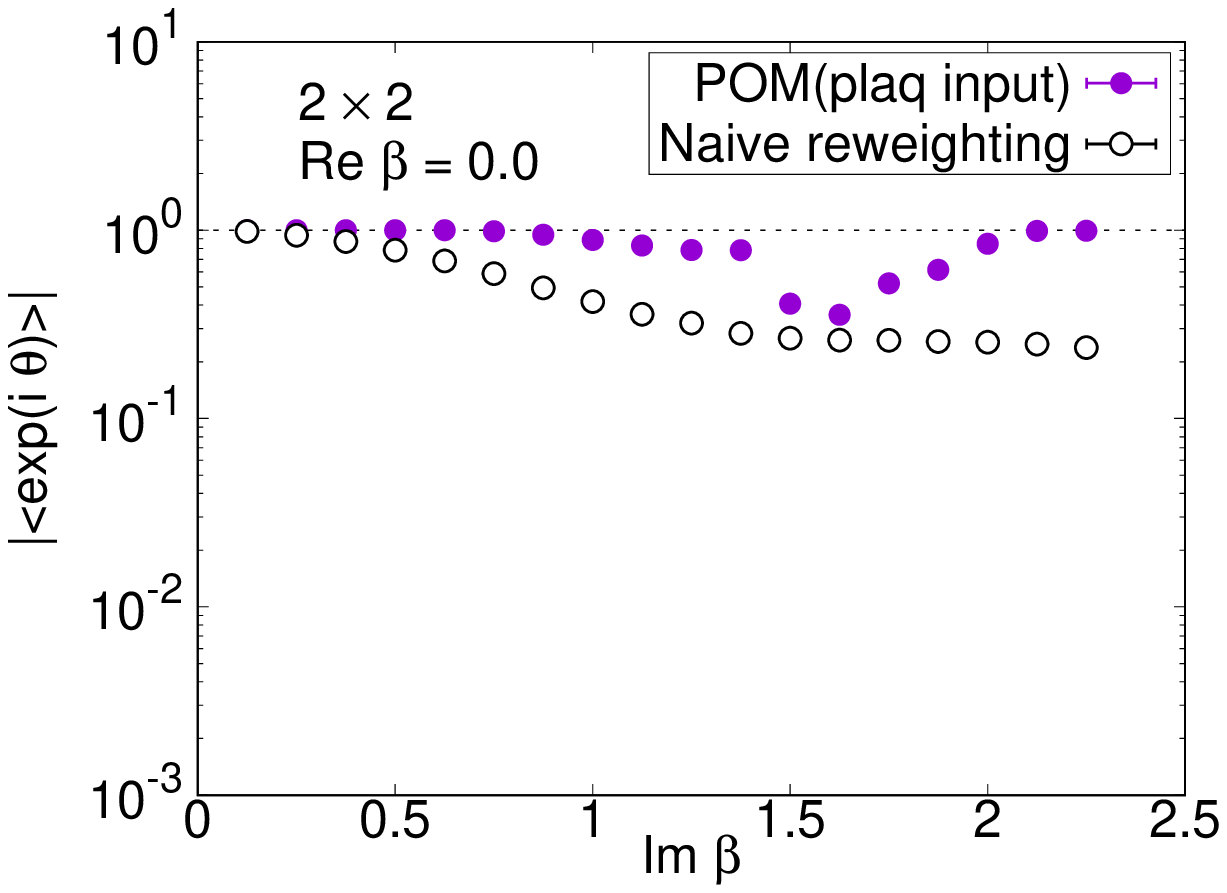}
 \includegraphics[width=7.5cm]{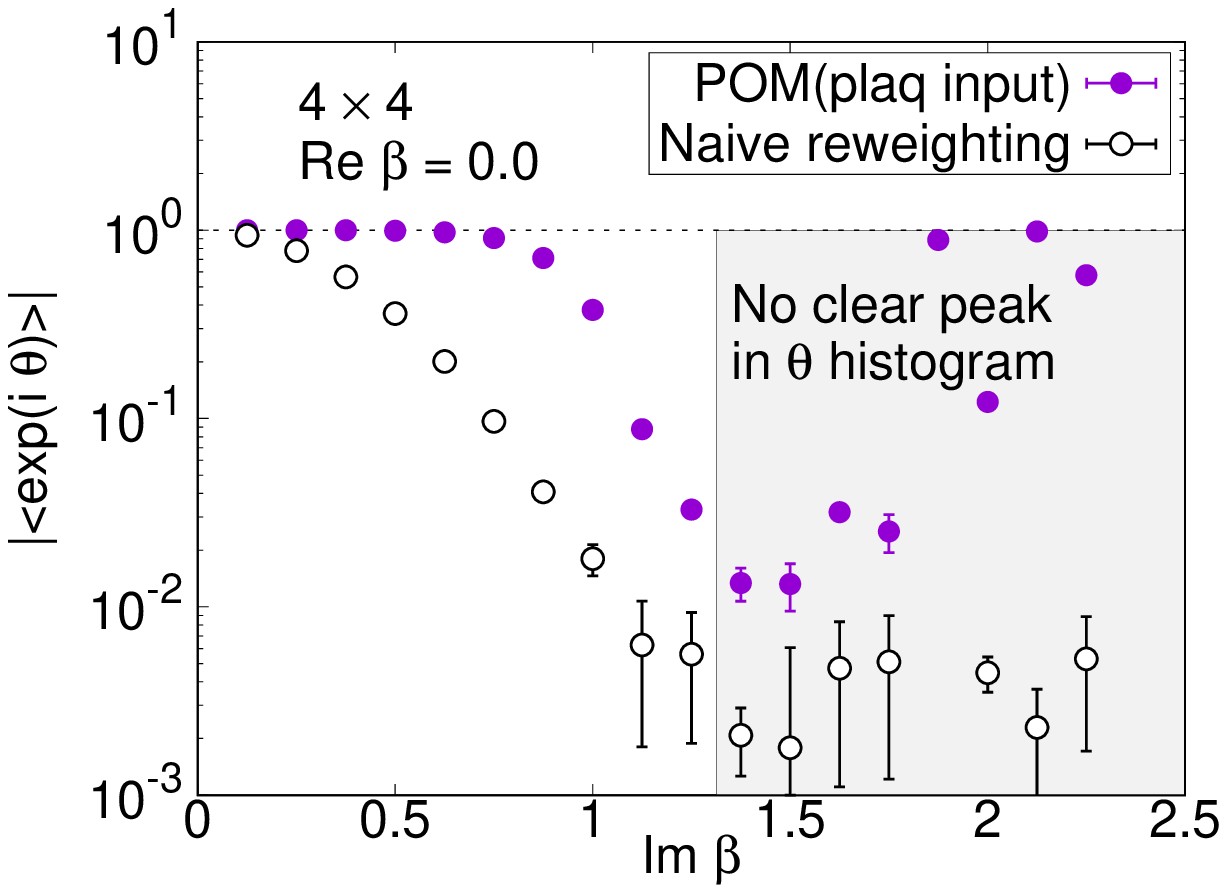}
 \includegraphics[width=7.5cm]{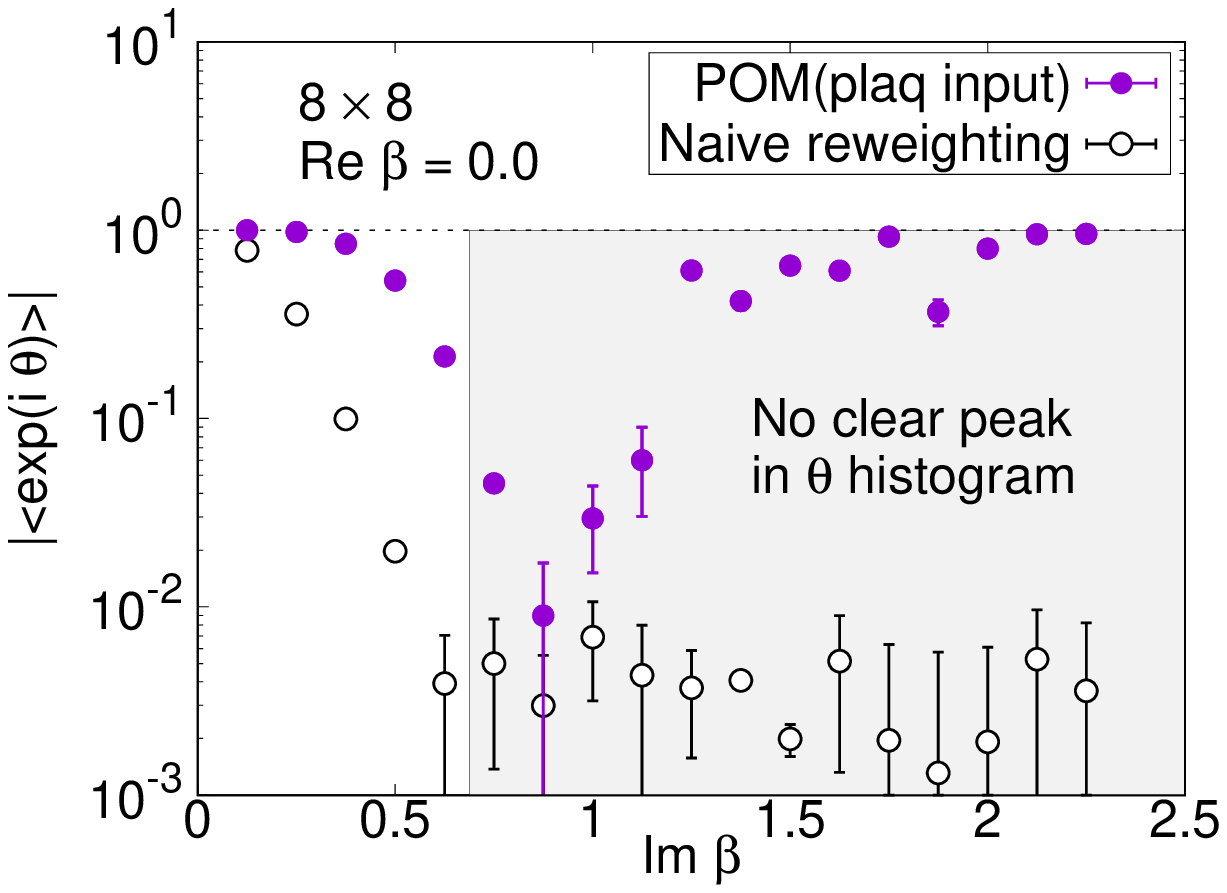}
 \caption{The average phase factors with and without the path optimization at $\beta = 0.0 + (0.25$--$2.25) i$ on $2 \times 2$ (upper panel), $4 \times 4$ (middle panel), and $8 \times 8$ (lower panel) lattices.}
 \label{Fig:beta_i-apf_beta}
\end{figure}

\begin{figure}[ht]
 \centering
 \includegraphics[width=7.5cm]{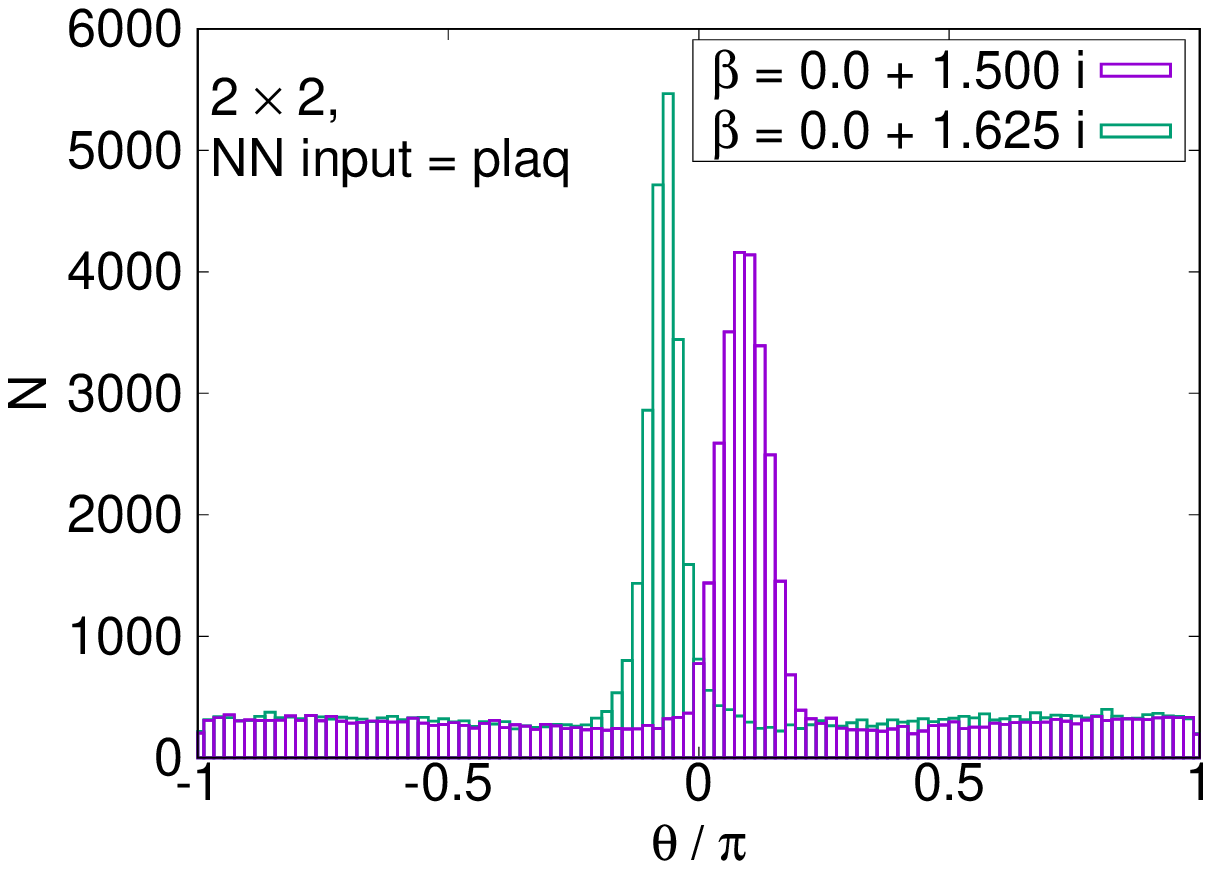}
 \includegraphics[width=7.5cm]{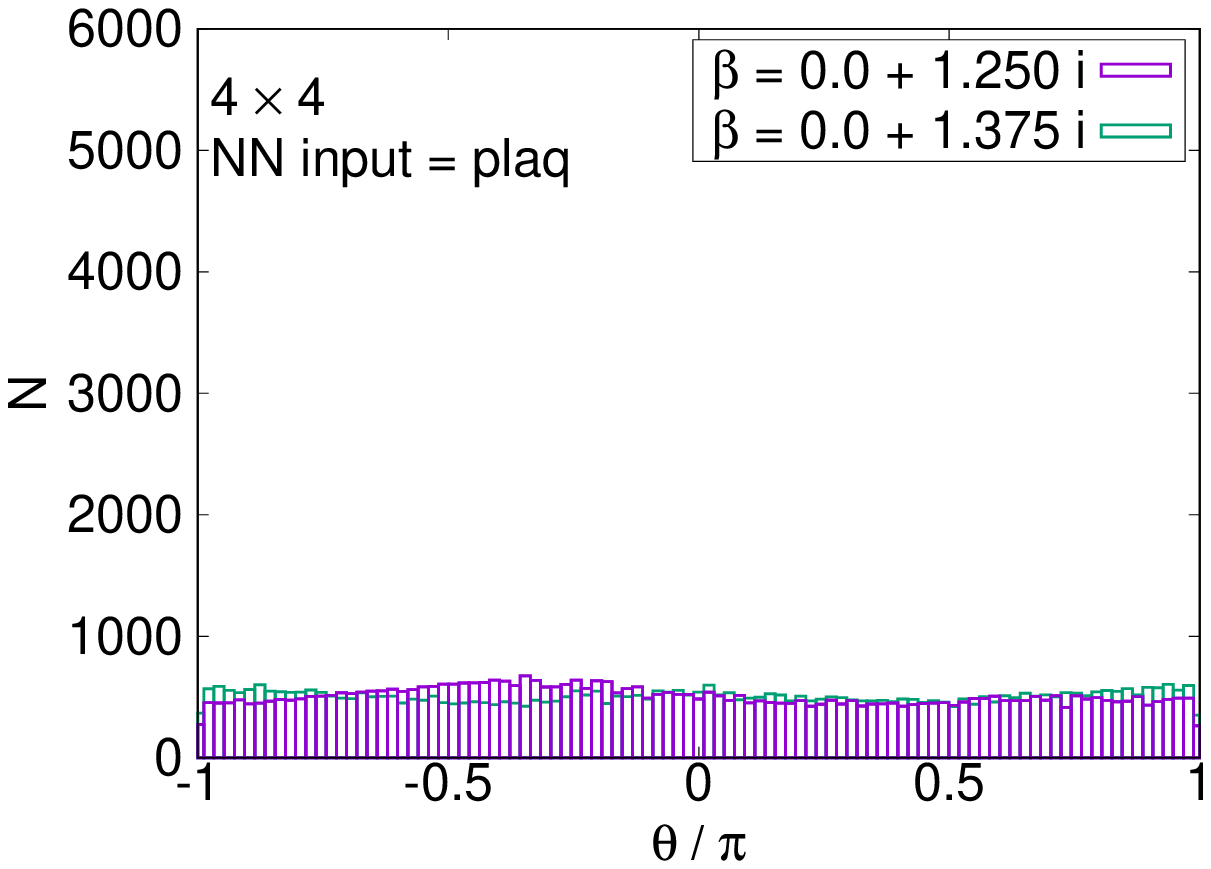}
 \includegraphics[width=7.5cm]{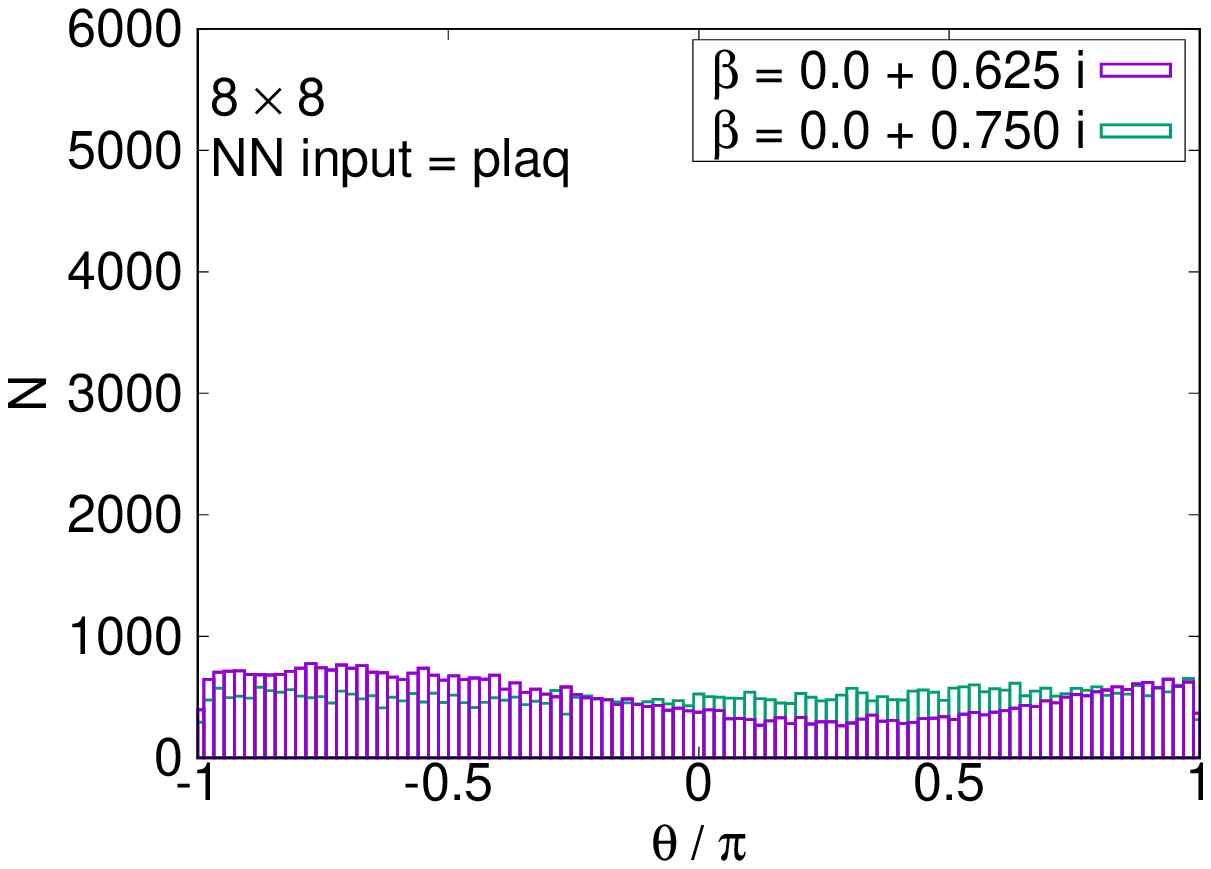}
 \caption{Histogram of the phases with the path optimization around $\beta_c$ on $2 \times 2$ (upper panel), $4 \times 4$ (middle panel), and $8 \times 8$ (lower panel) lattices.}
 \label{Fig:histogram_beta_c}
\end{figure}

\begin{figure}[ht]
 \centering
 \includegraphics[width=7.5cm]{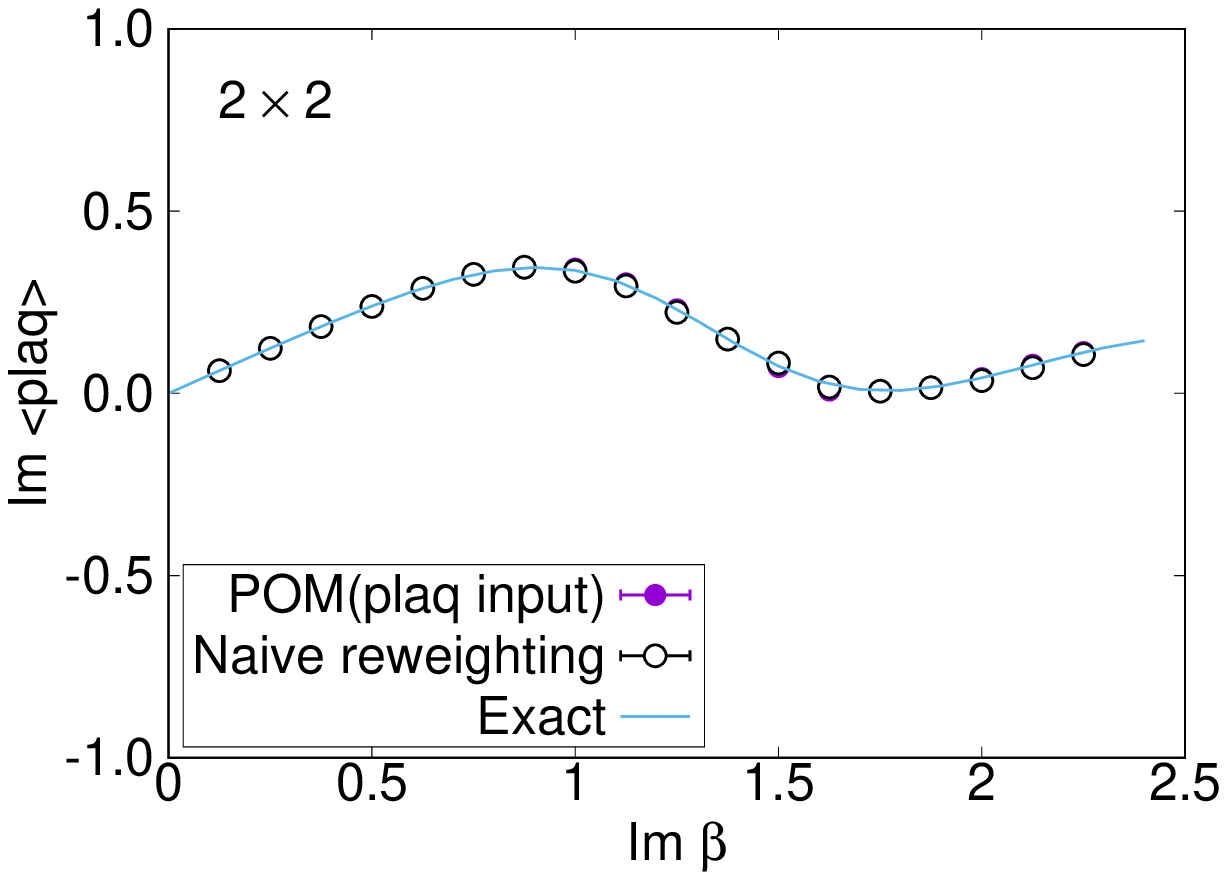}
 \includegraphics[width=7.5cm]{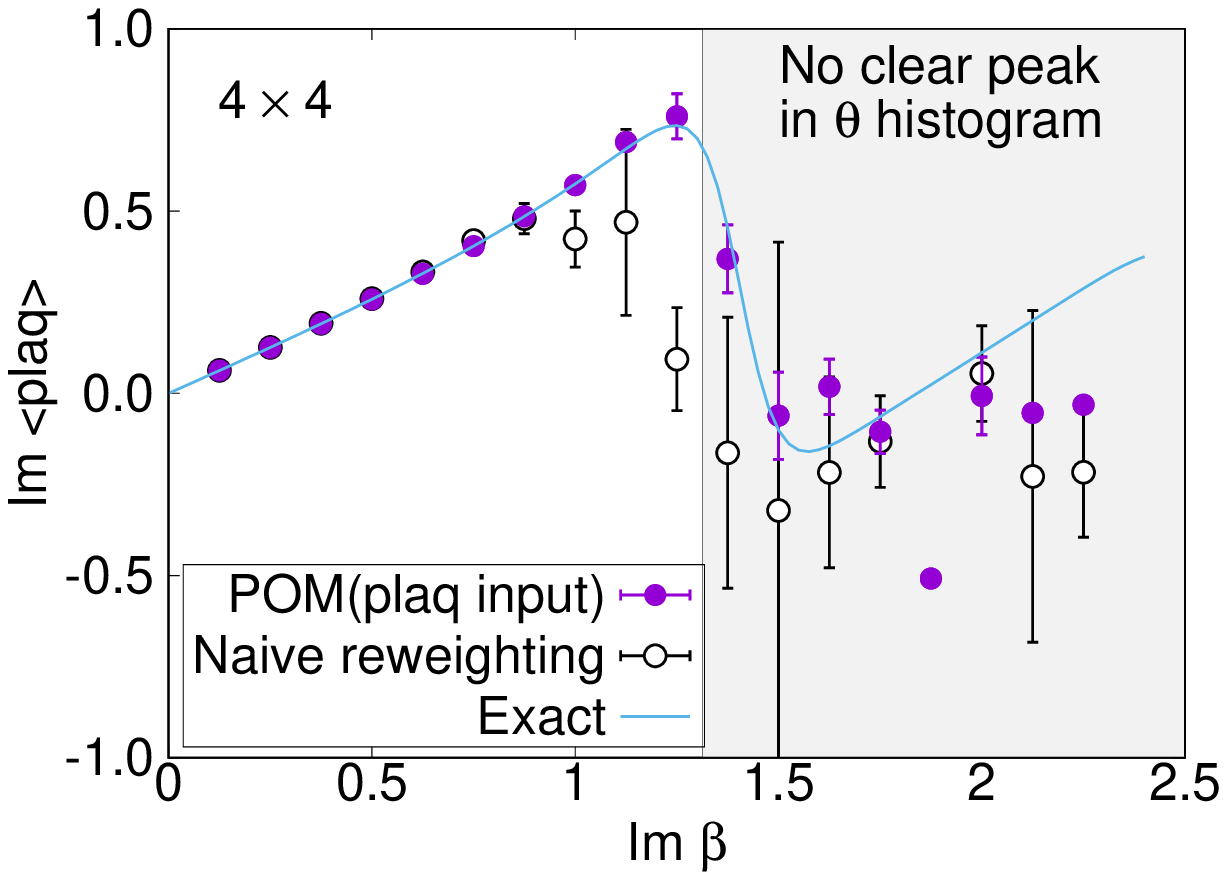}
 \includegraphics[width=7.5cm]{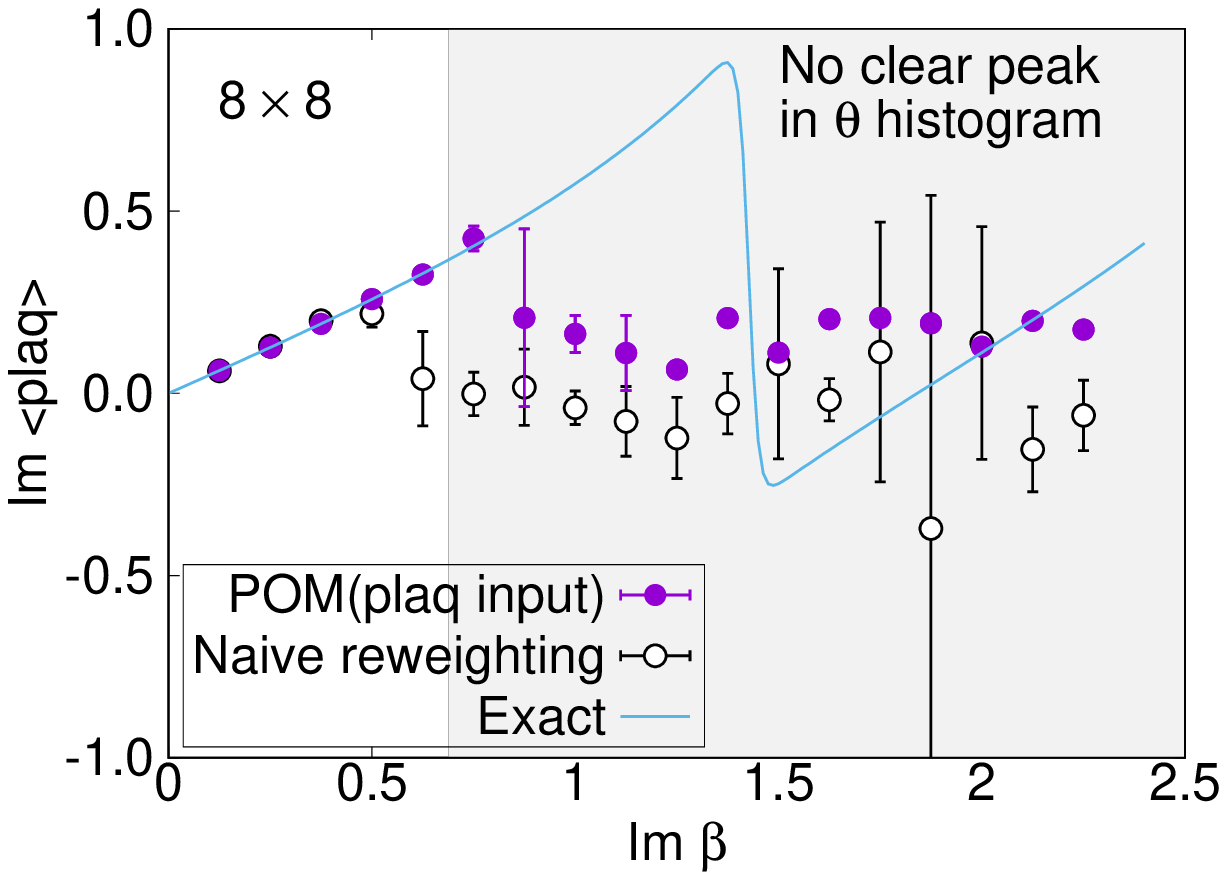}
 \caption{Expectation values of the imaginary part of plaquette with and without the path optimization at $\beta = 0.0 + (0.25$--$2.25) i$ on $2 \times 2$ (upper panel), $4 \times 4$ (middle panel), and $8 \times 8$ (lower panel) lattices.}
 \label{Fig:beta_i-plaq}
\end{figure}

\begin{figure}[t]
 \centering
 \includegraphics[width=7.5cm]{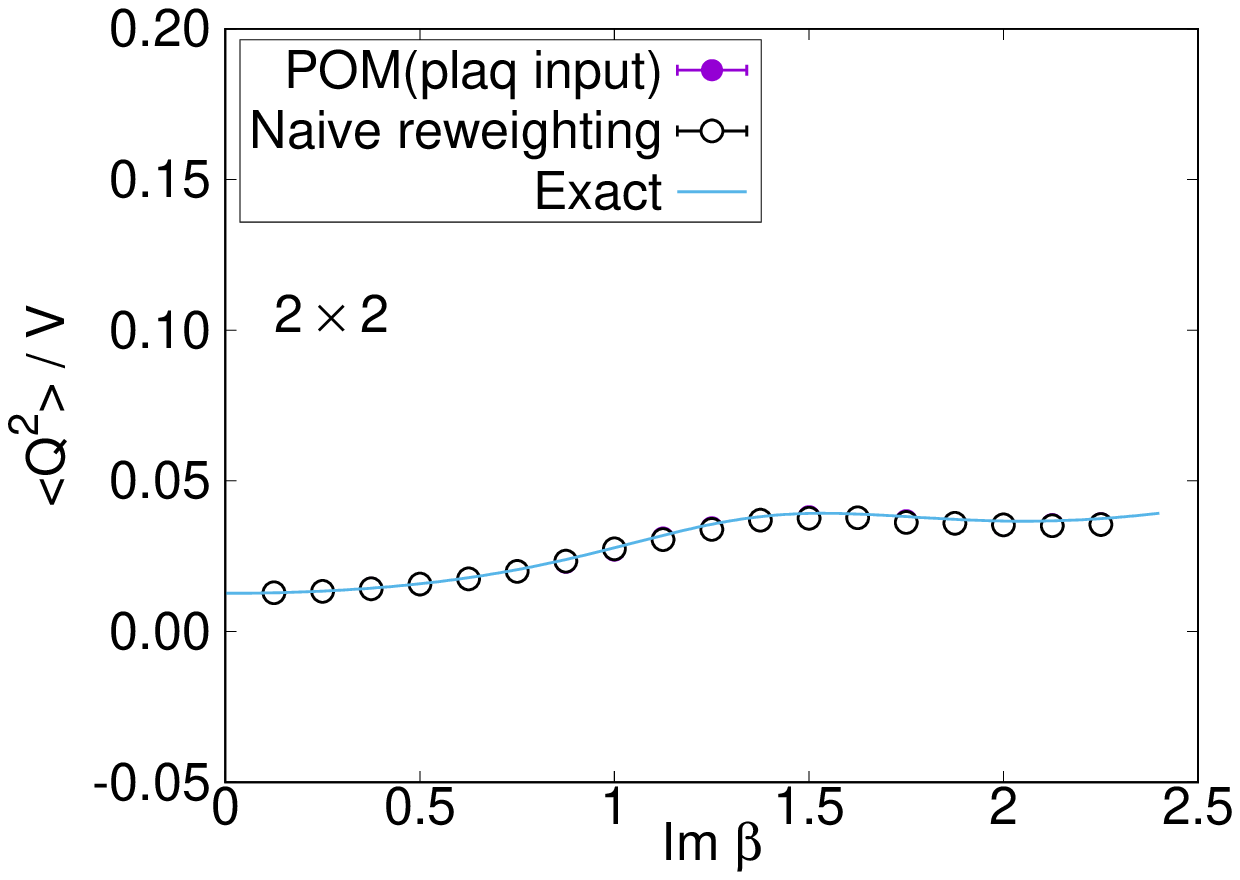}
 \includegraphics[width=7.5cm]{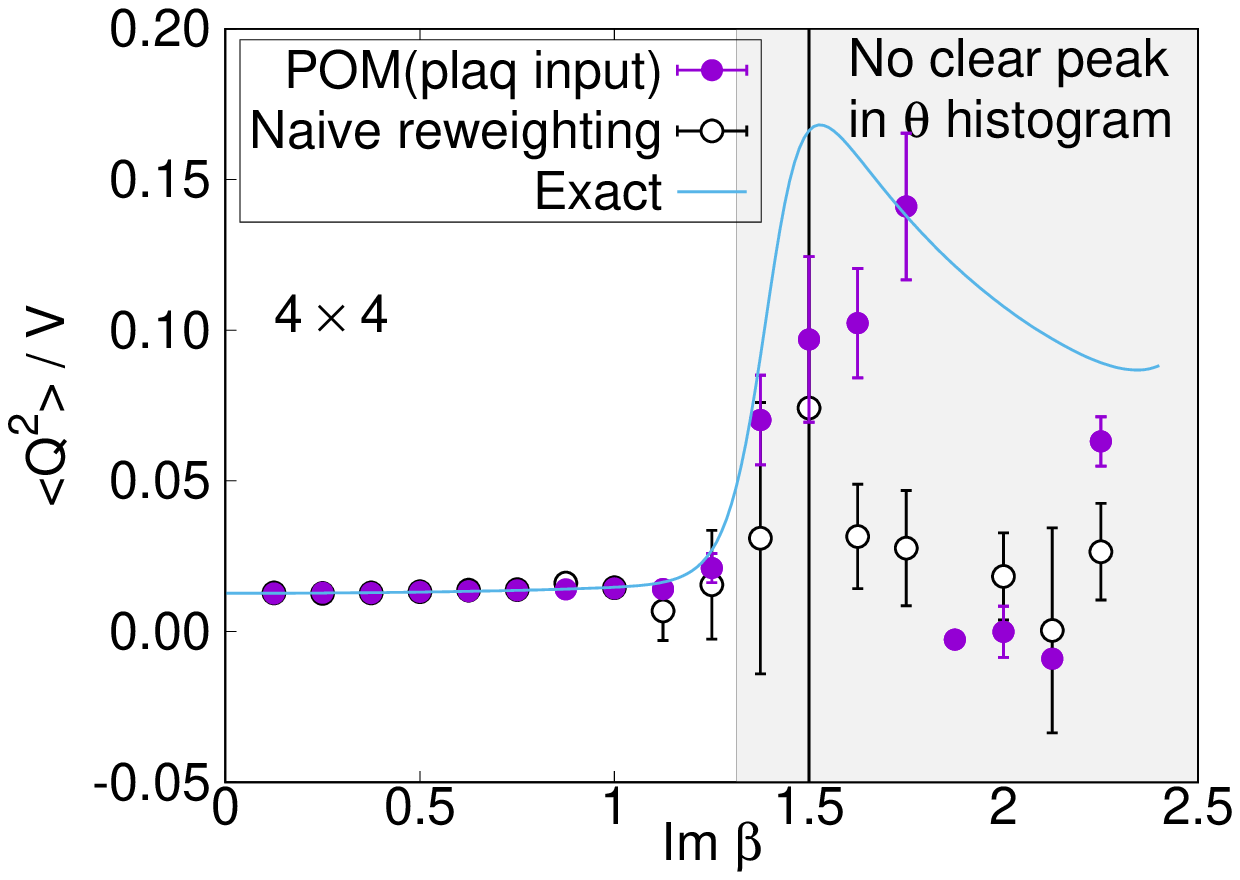}
 \includegraphics[width=7.5cm]{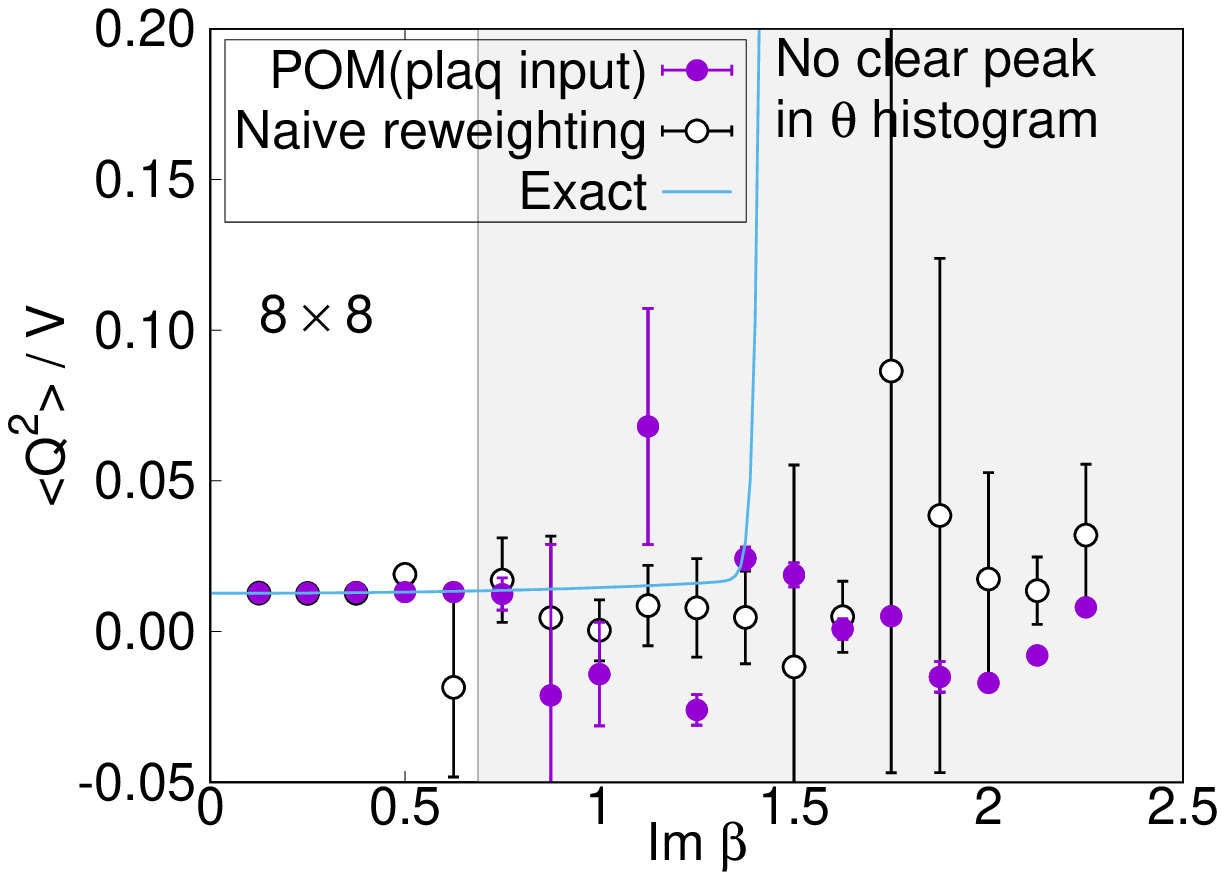}
 \caption{Topological charge susceptibility with and without the path optimization at $\beta = 0.0 + (0.25$--$2.25) i$ on $2 \times 2$ (upper panel), $4 \times 4$ (middle panel), and $8 \times 8$ (lower panel) lattices.}
 \label{Fig:beta_i-Q2}
\end{figure}

\section{Numerical setup and result}
\label{Sec:Numerical}

\subsection{Setup}

We evaluate performance of the POM for $\beta = 0.0 + (0.25$--$2.25) i$.
The sign problem is originated from the imaginary part of $\beta$.
We generate gauge configurations by the Hybrid Monte-Carlo algorithm in the POM.
The total number of configurations is $50000$.
Statistical errors are estimated by the Jackknife method with the bin size of $250$.
The neural network utilizes ADADELTA optimizer~\cite{Zeiler2012:adadelta} combined with the Xavier initialization~\cite{Glorot2010:understanding}.
The parameters in Eqs.\eqref{Eq:FNN1} and \eqref{Eq:FNN2} are optimized during learning.
The flow chart is displayed in Ref.~\cite{Kashiwa:2020brj}.
We set the learning rate to 1 and the decay constant 0.95, combined with the batch size of 10.
The number of units in the input layer reflects our choice of $t$ in Eqs.~\eqref{Eq:t_P}, \eqref{Eq:t_U} as
\begin{align}
 ({\rm i}) \,  &N_\mathrm{input}^{\rm plaq} = 2 \, n_\mathrm{deg}^{\rm plaq} \mbox{ for plaquette input},
 \\
 ({\rm ii}) \, &N_\mathrm{input}^{\rm link} = 2 \, n_\mathrm{deg}^{\rm link} \mbox{ for link input},
\end{align}
where $n_\mathrm{deg}^{\rm link} = 2 N_1 N_2$ and $n_\mathrm{deg}^{\rm plaq} = N_1 N_2$.
The number of units in the output layer is common to (i) and (ii), $N_\mathrm{output} = n_\mathrm{deg}^{\rm link}$.
We employ a single hidden layer with $N_\mathrm{hidden} = 10$ hidden units on $2 \times 2$ lattice, $16$ on $4 \times 4$ lattice, and $64$ on $8 \times 8$ lattice, respectively.
As we set the $N_\mathrm{hidden}$ proportional to the volume, the cost of the neural network is $O(n_{\rm deg}^2)$ and is $O(n_{\rm deg}^3)$ for the Jacobian.

\subsection{Result}

Figure~\ref{Fig:iter-apf_beta_i} exhibits the neural-network{\textendash}step-number dependence of the exponential moving average of the average phase factor
$\left\langle \exp(i \theta) \right\rangle_{\rm EMA}$
at $\beta = 0.0 + 0.5 i$ on a $4 \times 4$ lattice as a typical example.
The path optimization with the plaquette input successfully enhances $\left\langle \exp(i \theta) \right\rangle_{\rm EMA}$, while that with the link variable input does not.
Similar behavior is also observed at other values of $\beta$ on $2 \times 2, 4 \times 4$ and $8 \times 8$ lattices.
Our result verifies the advantage of the gauge invariant input to the neural network for the 2-dimensional $U(1)$ gauge theory with the complex coupling.

The enhancement with the plaquette input is confirmed in the histogram of the phases, shown in Fig.~\ref{Fig:histogram_beta0.5}.
While the naive reweighting gives a broad distribution of the phase factor, the path optimization significantly sharpens the peak structure.
We stress that the POM works even on the $8 \times 8$ lattice, where the sign problem is severer.
Although the naive reweighting has almost flat dependence on the phase, the POM can still extract a peak structure around $\theta / \pi \sim -0.2$.

Figure~\ref{Fig:L2-apf_beta} represents the volume dependence of the average phase factor at $\beta = 0.0 + 0.5 i$ with additional simulation results on $6 \times 6$, $10 \times 10$ and $12 \times 12$ lattices.
The naive reweighting leads to steep exponential fall-off as a function of the volume.
The sign problem becomes extremely severer toward the infinite volume limit.
The path optimization evidently changes the volume dependence of the average phase factor to be milder.
It indicates better control of the sign problem.

We plot the average phase factors with and without the path optimization as functions of $\mathrm{Im}\,\beta$ in Fig.~\ref{Fig:beta_i-apf_beta}.
The average phase factors are enhanced by factors of up to 4 on $2 \times 2$, 21 on $4 \times 4$, and 27 on $8 \times 8$ lattices, respectively.
The enhancement decreases, however, as we approach the critical coupling ${\rm Im } \beta_c \sim 1.5$ where the partition function becomes zero, corresponding to the Lee-Yang zero.
While clear peaks are still visible in the histogram of the phases by the path optimization even at $\beta \sim \beta_c$ on $2 \times 2$ lattice, no clear peak is obtained and the neural network eventually becomes unstable around $\beta_c$ on $4 \times 4$ and $8 \times 8$ lattices, as displayed in Fig.~\ref{Fig:histogram_beta_c}.
We need further improvement of the POM around $\beta_c$ on large lattices.

Comparison with the exact solution~\eqref{Eq:exact} is accomplished for the expectation values of the plaquette and the topological charge.
The result for the plaquette is plotted in Fig.~\ref{Fig:beta_i-plaq} and for the topological charge in Fig.~\ref{Fig:beta_i-Q2}.
The naive reweighting works in a small $\beta$ region but starts to deviate from the exact value with uncontrolled errors as $\beta$ becomes larger.
Our data using the POM agree with the exact solution, as long as we find a peak in the histogram of the phases.
The valid region of the POM is definitely extended to larger $\beta$.
Deviations from the exact solution are also observed by the path optimization case, if we have no clear peak in the histogram of the phases, where enhancement of the phase factor by the path optimization is still limited to be small or the path optimization is unstable.
One reason of the failure is a restriction of the statistics.
The machine learning requires more data to find the best path especially for a system with a large degrees of freedom.
Another possibility is the effect of the multimodality.
Though there can be several regions contributing to the integral (relevant thimbles),
the optimized path may emphasize only a part of them.
Quantitative evaluation of the systematic errors of the POM result is still difficult and is beyond the scope of this paper.
It is an important future work.
Nevertheless, these results demonstrate the superiority of the POM over the naive reweighting for the analysis of the $U(1)$ gauge theory with the complex $\beta$.

\section{Summary}
\label{Sec:Summary}

We established the efficiency of gauge invariant input in the POM for the 2-dimensional $U(1)$ gauge theory with a complex coupling.
While the path optimization using the link variable input without gauge fixing shows no gain, the optimization with the gauge invariant input shows a clear increase in the average phase factor through the neural network process.
The gain is up to 4 on $2 \times 2$, 21 on $4 \times 4$, and 27 on $8 \times 8$ lattices.
Even on $8 \times 8$ lattice at $\beta = 0.0 + 0.5 i$, the POM using the gauge invariant input successfully identifies a peak structure in the histogram of the phases, where the naive reweighting shows no peak in the histogram.
We confirm the volume dependence of the average phase factor becomes much milder by the POM with the gauge invariant input.
We also calculated expectation values of the plaquette and the topological charge for comparison with their exact solutions.
Our results agree with the analytical values, as long as we find a peak in the histogram of the phases.
The valid region is clearly enlarged to larger $\beta$, compared with that by the naive reweighting method.
It is encouraging toward the POM analysis in more realistic cases.

In the severer sign problem region near the critical point on the large volume, the gain of the POM is less clear.
Further improvement of the POM is required.
One possible direction is the incorporation of larger Wilson loops and the Polyakov lines as input to the neural network.
Another direction is the adoption of novel approaches which respect the gauge symmetry in the neural network, such as the lattice gauge equivariant convolutional neural networks~\cite{Favoni:2020reg} and the gauge covariant neural network~\cite{Tomiya:2021ywc}.
In addition, there is a different approach that modifies the action instead of the integral path~\cite{Tsutsui:2015tua,Doi:2017gmk,Lawrence:2020kyw}.
The combination of the modifications of the action and the integral path seems to be interesting.

\begin{acknowledgments}
Our code is in part based on LTKf90~\cite{Choe:2002pu}.
This work is supported by JSPS KAKENHI Grant Numbers 
JP18K03618, JP19H01898, and JP21K03553.
\end{acknowledgments}

\bibliography{ref.bib}

\end{document}